\documentclass{article}
\usepackage[utf8]{inputenc}
\usepackage{titling}
\thanksmarkseries{alph}
\usepackage{float}
\usepackage{graphicx}
\usepackage[font=scriptsize,labelformat=empty]{subcaption}
\usepackage{authblk}

\title{The perils of automated fitting of datasets: the case of a wind turbine cost model}
\author[1]{Claude Kl\"ockl \thanks{claude.kloeckl@boku.ac.at}}
\author[1]{Katharina Gruber\thanks{katharina.gruber@boku.ac.at}}
\author[1]{Peter Regner\thanks{peter.regner@boku.ac.at}}
\author[1]{Sebastian Wehrle\thanks{sebastian.wehrle@boku.ac.at}}
\author[1]{Johannes Schmidt\thanks{johannes.schmidt@boku.ac.at}}

\affil[1]{
Institute for Sustainable Economic Development, University of Natural Resources and Life Sciences, Feistmantelstraße 4,
1180 Vienna, Austria
}

\usepackage{amsmath, amssymb, amsfonts, mathrsfs}
\usepackage{hyperref}
\usepackage{graphicx}
\begin{document}  

\maketitle

\section{Introduction}

Planning of future electricity systems is mostly performed with optimization models, which allow deriving optimal portfolios of generation technologies to e.g. minimize $\mathrm{CO}_2$ emissions. A highly relevant input parameter for electricity system models are technology specific generation costs, and here in particular of variable renewable energies such as solar photovoltaics and wind, as those sources will significantly increase their share in electricity generation. In contrast to fuel-based energy sources, the costs of variables renewable energies and in particular wind energy are almost entirely determined by their investment cost. Unfortunately, wind turbine investment costs are hardly ever made public by the involved companies, making this modelling prerequisite not directly available to energy planners.
%However, many countries release detailed lists of all their installed turbines including physical characteristics (i.e. height, diameter and rated power).
Openly published costs estimation methods, that estimate turbine investment costs from publicly available physical characteristics are therefore very helpful for energy system modellers. Rinne et al. \cite{Rinne2018} have published such a model, which could be of high use for the modelling community. However, we show analytically and by example that the model uses an investment cost model that exhibits unrealistic scaling behaviour in certain, empirically relevant, parameter regimes. The model is based on a linear regression which will perform reasonably well for data that is similar to its training data. However, we demonstrate that the model exhibits a strong generalization error (i.e. error on data that is unsimilar to its training data). Typically, generalization errors indicate that a model is not well-representative of the fundamental dynamics governing the modeled system.
We recommend a set of axioms for investment cost models, that we believe enforces a stricter adherence of cost models to how costs arise from wind turbines.

\section{The Rinne et al. model of wind turbine cost}

The functional form of the model is described by Equation (8) in \cite{Rinne2018}, which we restate as

\begin{equation}
\label{RinneImplicit}
y = \beta_{1} f_{1}(x_{1}) + \beta_{2} f_{2}(x_{2}) + \beta_{3} f_{3}(x_{3}) + C.
\end{equation}

The model is a linear combination of several basis functions $f_{i}$, specifically $x,x^{2}, ln(x)$ and $\sqrt{x}$. All possible combinations of $f_{i}$ are considered as candidate models. Subsequently, linear regressions are run for all models and the one with the lowest RMSE is selected. In accordance with the final wind turbine cost model presented in \cite{Rinne2018}, we write (8) in its explicit form

\begin{equation}
\label{RinneExplicit}
\mathrm{SpecificCost}(hh,p,r,age) = 620 \mathrm{ln}(hh) - 1.68 \frac{p}{r^2 \pi} + 182 \sqrt{age} - 1005.
\end{equation}

We inserted all constants as in \cite{Rinne2018} and renamed the parameters to easier identify their meaning.
Specific costs denote costs per rated capacity.
Parameter $hh$ is the hub height, $p$ is rated capacity, $r$ is the rotor radius and $age$ indicates when the turbine initially was introduced  to the market. Note that we employ four parameters instead of three, since we replaced the composite input parameter specific power $\frac{p}{r^{2}\pi}$ from \cite{Rinne2018} with its elementary constituents $p$ and $r$. Analytically, these models are identical. 

\section{Where the model fails}

Equation \eqref{RinneExplicit} implies that a wind turbine's total costs are

\begin{equation}
\label{RinneTotalCost}
\mathrm{TotalCost}(\mathrm{hh},p,r,\mathrm{age}) = p (620\mathrm{ln}(\mathrm{hh})-1.68 \frac{p}{r^2\pi}+182\sqrt{\mathrm{age}}-1005)
\end{equation}

\begin{figure}
  \caption{Comparison of 3 turbines estimated investment costs.}
\label{fig:1}
  \includegraphics[width=1\linewidth]{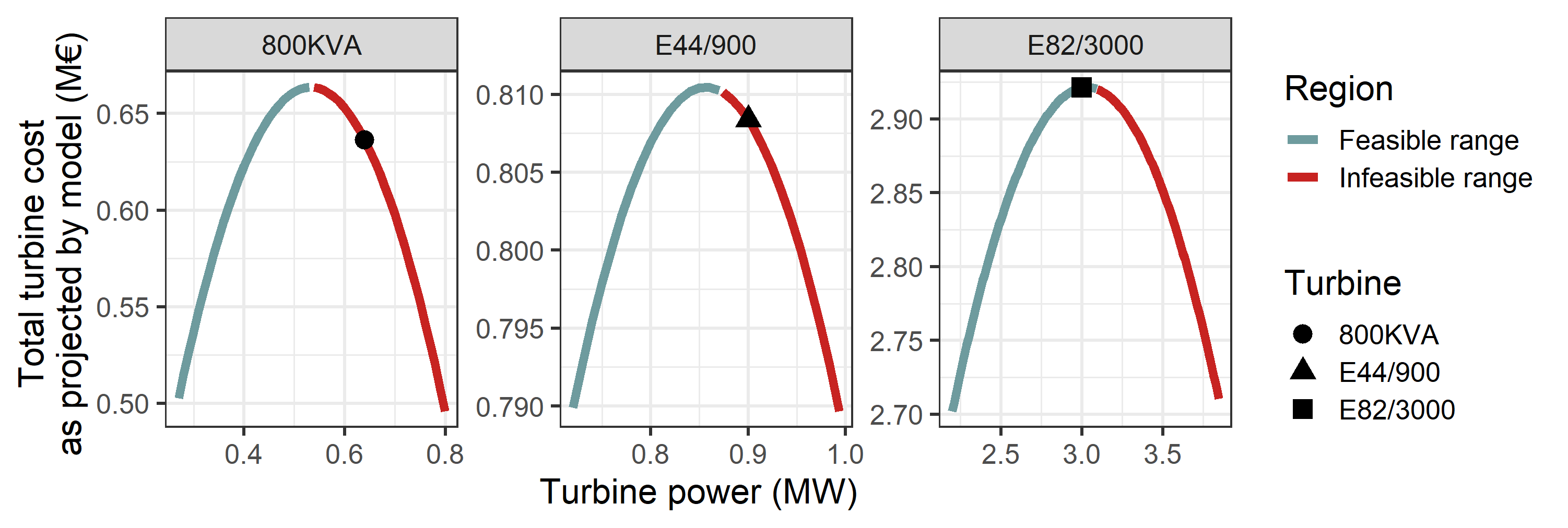}
  \subcaption{The 800KVA and E-44 lie in the implausible range, while the E-88 lies barely in the plausible range. The 800KVA is a historic turbine. It illustrates that the model does not generalize well for some turbines. The E44 and E82 are currently sold by ENERCON, one of the largest German turbine producers. E44 shows that the model produces costs in the implausible range also for modern turbines. E82's larger diameter is more representative of today's average diameter. It shows, that the model predictiction of costs for a standard turbines are not strictly analytically wrong, but in a very flat region of the cost curve.}
\end{figure}

Our main concern is the scaling behavior of (\ref{RinneTotalCost}). In essence, a consistent model should mirror a realistic scaling behavior of all parameters. In particular, total costs of a large turbine should be predicted higher by the model than costs of a smaller one, everything else being equal, although specific costs may decrease with size. Figure \ref{fig:1} shows that basic scaling relations of the rated power $p$ are violated in Equation \eqref{RinneTotalCost} for certain parameter regimes. It depicts the scaling of investment cost with respect to rated power $p$ and fixed $r$, $age$, and $hh$ for the  turbine models Best-Romani 800KVA, Enercon E-44/800 and Enercon E-82/3000 respectively (see Figure \ref{fig:1}, turbine data from \cite{windpower.net}).

Note that Equation \eqref{RinneTotalCost} has a quadratic term in $p$ with a negative coefficient, therefore the function has one maximum. For $p$ above that maximum, total turbine costs decrease in absolute terms while increasing rated capacity $p$. This is counter-intuitive and cannot be explained by economies of scale, which should at best lead to declining marginal increases in total cost.

We refer to $p$ values below the maximum cost as plausible region (shown in red) and above the extremum as implausible (shown in turquoise). Here, plausible means that a point on the curve may potentially replicate real cost behaviour, even though we make no judgement on the correctness of the result in this regime. 

One could presume that any reasonable choice of turbine parameters remains in the plausible region. This, however, is not the case as the examples of the  Best-Romani 800KVA and the Enercon E-44/800 lie in the implausible region in Figure \ref{fig:1} \cite{windpower.net}.
Points in the implausible region seem incorrect a priori due to an implausible scaling behaviour. In that region, increasing turbine capacity, everything else being equal, will decrease total investment costs predicted by the cost model. For some values, turbine investment costs even become negative. 
Consequently, both cost estimates are lower for that turbine than for a similar turbine with exactly the same specification, but lower power rating. We remark that while the Best-Romani is a very old model, the E-44 is actively used and currently sold by Enercon, one of the largest German wind power producers.

We have included in Figure \ref{fig:1} the Enercon E-88/3000 as an additional example of a turbine that lies in the plausible region, but is close to the border of feasibility.
We note that our analysis can not directly point at wrong scaling behavior within the plausible region and hence that costs of the E-88 are not demonstrably incorrect. 
Nonetheless, we believe that a wind turbine cost function should not have a maximum when scaling up physical parameters such as rated power. Hence, the very flat increases in costs around the maximum are likely an artifact of the maximums wrong existence and also the correctness of such borderline cases as the E-88 has to be questioned.
We were able to find 12 turbine models that reside within 20\% of the maximum, and whose cost estimates should be used with caution therefore\footnote{Namely: Best-Romani 800KVA,
Enercon E44/900,
Enercon E82/3000,
Storc NEWECS-25,
XEMC-Darwind XD115-5000,
Windflow 45-500,
Enercon E70/2300,
Elsam Tjaereborg,
Enercon E82/2350,
United Energies UE 1.65 70 IA,
Bard 6.5,
Storc NEWECS-45 (see \cite{windpower.net})
}.

\begin{figure}
  \caption{Histogram of turbine models' predicted marginal costs for 1kW extra capacity}
\label{fig:2}
  \includegraphics[width=1\linewidth]{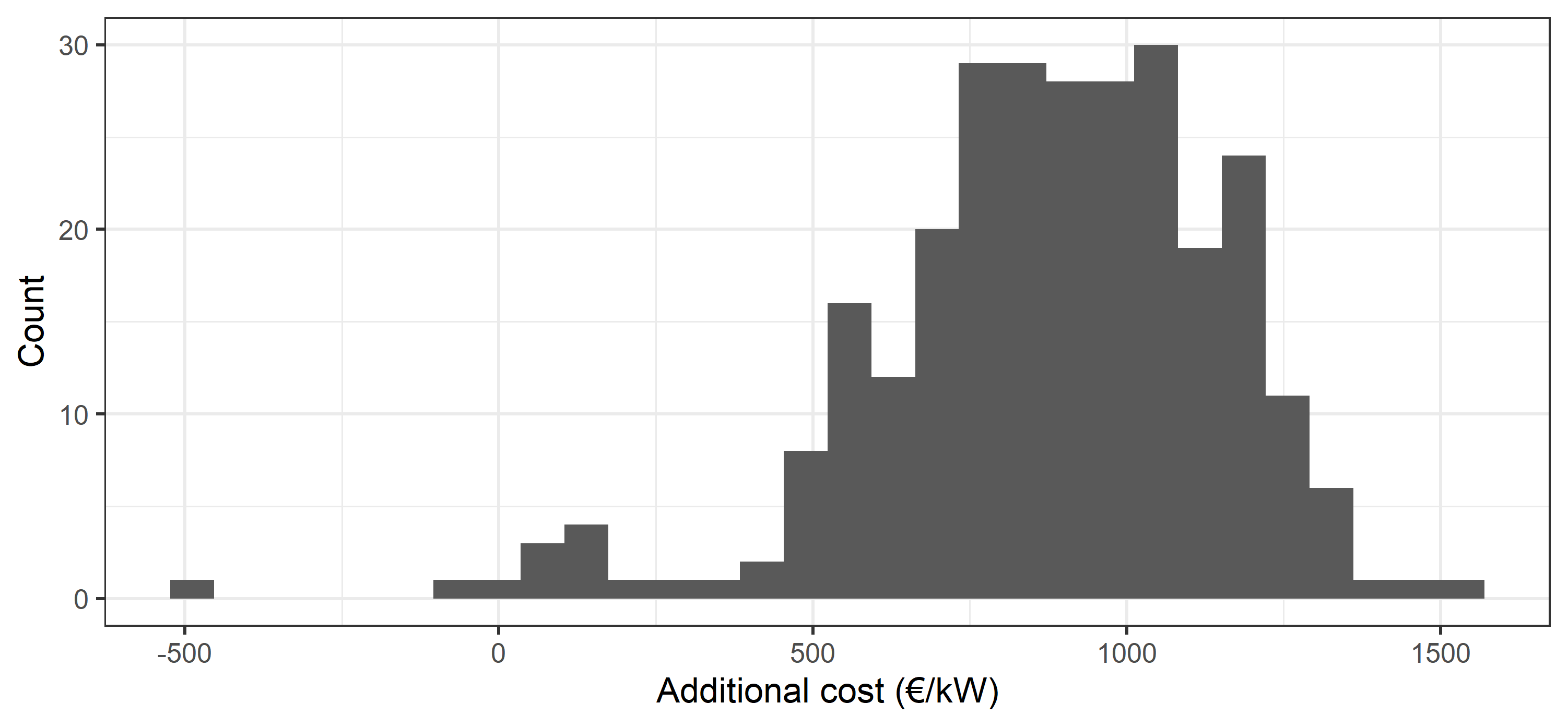}
  %\subcaption{B}
\end{figure}

Figure \ref{fig:2} shows how many turbines in our dataset are affected by questionable predictions, indicating the marginal costs of increasing a turbines generator by 1kW for all turbine types contained in \cite{windpower.net} as predicted by Equation \eqref{RinneExplicit}.
It is evident that extreme cases as $-500\$/kW$ or $0\$/kW$ marginal costs have to be wrong as a larger generator will at least incur some additional cost. Moreover, the long left tail of the distribution is a further indication that the model fails for some relevant parameters. For instance 8\% of all turbines in \cite{windpower.net} fall below a very low marginal cost of 500\$/kW.

\section{Why the model fails}

Here we discuss why the model fails in delivering valid results over the whole parameter range. In particular this is related to the choice of (i) variables and (ii) basis functions in the model.

\subsection{Choice of variables in the model}
The selection of variables in \cite{Rinne2018} is neither motivated nor validated. We understand that hub height and age as a proxy of technological progress may have an impact on costs. However, the parameter specific power is not necessarily directly related to investment costs. While the authors convincingly argue that the specific power of recent turbine models has been decreasing over time and cite related work that shows that decreased specific power leads to decreased levelized costs of electricity (LCOE), that does not directly indicate that a lower specific power reduces investment costs.\\
Observe that specific power is a ratio of two parameters. Thus, increasing the specific power can be attained in two different ways: either by increasing the size of the generator or decreasing the length of blades. While smaller blades most likely lead to lower costs, a higher rated capacity most likely increases costs. Therefore, an increase in specific power may have opposite effects on costs. Figure \ref{fig:1} shows that this is empirically relevant.

\subsection{Choice of fitting functions}
 The authors don't motivate why the particular basis functions $x,x^{2},ln(x),\sqrt{x}$ were chosen. First, it is unclear why polynomials of other degrees, the exponential function or any other functions are disregarded. Others have employed different functional forms. For instance, the National Renewable Energy Laboratory (NREL) cost model \cite{NREL2006} uses linear combinations of polynomials with fractional exponents between 2 and 3 to model realistic scaling behaviour. Second, the chosen functional form  disallows some predictions: the choice to model age (in their definition years before 2016) by a squareroot makes it impossible to predict costs for turbines built after 2016, due to negative terms under the squareroot.
 An additional problem is model selection. Picking a model through minimizing RMSE is dependent on the allowed basis functions. It is well known that $n+1$ data points can be interpolated exactly by a degree $n$ polynomial. Therefore a polynomial model of sufficiently high degree is guaranteed to attain RMSE 0.
 Model selection should therefore not be performed solely via the RMSE but by requiring simple axiomatic minimal requirements from the model.\\ Notably, we propose to require \textit{physical monotonicity} i.e. any turbine cost model to be a monotonically increasing function in all physical dimensions (i.e. height, diameter) but also in terms of electrotechnical dimensioning (i.e. rated capacity).
Monotonicity rules out some possible functional forms such as ratios or negatively squared terms involving the aforementioned parameters.
It is however not sufficient to completely determine a functional form of wind turbine costs in general.
We hence propose \textit{costs-to-scale} assumptions on increasing or decreasing return-of-scale for costs, that might further narrow down the functional forms by ruling out for instance either logarithmic or exponential forms.
For instance the aforementioned NREL model \cite{NREL2006} is inline with both monotonicity and increasing costs of scale and would hence be acceptable under our proposed axioms.

\section{Conclusions}

Rinne et al. \cite{Rinne2018} conduct a detailed analysis of the impact of wind turbine technology and land-use on wind power potentials, which allows important insights into each factor's contribution to overall potentials. The paper presents a detailed and very valuable model of site-specific wind turbine investment cost (i.e. road- and grid access costs), complemented by a model used to estimate site-independent costs.  
However, the site-independent cost model is flawed in our opinion. This flaw most likely does not impact the results on cost supply-curves of wind power presented in the paper. However, we expect a considerable generalization error. Thus the application of the wind turbine cost model in other contexts may lead to unreasonable results. More generally, the derivation of the wind turbine cost model serves as an example of how applications of automated regression analysis can go wrong.

We therefore caution against a naive utilization in other studies. In particular, numerically solved optimization problems, where height, diameter and rated power are decision variables, may generate implausible results if relying on the model. The problem can be handled by (a) explicitly limiting the parameter space of the model to ensure validity of its results, (b) only applying the model to very common turbine designs or (c) replacing the model with a model that adheres to some basic axioms like assumptions on \textit{monotonicity} and \textit{returns-of-scales}. 
Such models exist already, for instance the NREL model \cite{NREL2006,Ryberg2018} satisfies our axioms.

Moreover, we hope that the article can serve as a starting point towards the axiomatization of wind turbine cost-models. We believe that such an axiomatization could increase the overall model quality in the field. The two minimal axioms we propose can not yet completely determine the functional form of a cost model, hence there remains considerable more work to be done toward a sound modelling of wind turbine investment costs.

\subsection{Acknowledgements}
Claude Kl\"ockl gratefully acknowledge financial support from the Anniversary Fund of the
Oesterreichische Nationalbank (OeNB), 18306.
Johannes Schmidt, Peter Regner, Katharina Gruber, Sebastian Wehrle and Claude Kl\"ockl thank the European Research
Council (‘reFUEL’ ERC-2017-STG 758149) for its financial support.
%\section{Competing interests}
%No competing interests to declare.

%\section{Author contributions}
%CK initially developed the idea, and wrote the first draft. KG, PR, and SW contributed with fundamental ideas and revised the paper extensively. JS contributed with the analysis of data and generating Figures 1 and 2.

\end{document}